\documentclass[aps,prb,twocolumn,amsmath,superscriptaddress]{revtex4-2}

\newcommand{\bea}{\begin{eqnarray}}
\newcommand{\eea}{\end{eqnarray}}
\newcommand{\beq}{\begin{equation}}
\newcommand{\eeq}{\end{equation}}

\newcommand{\AF}{AF}

\newcommand{\dd}{d^{\dagger}}
\newcommand{\dnd}{d^{\phantom{\dagger}}}

\usepackage{graphicx}
\usepackage{amssymb}
\usepackage{xcolor}
\definecolor{OliveGreen}{HTML}{3C8031}

\newcommand{\jbm}[1]{\textcolor{red}{#1}}

\begin{document}

\title{Charge and magnetic orders in a two-band model with long-range
  interactions for infinite-layer nickelates NdNiO$_2$}
\author{Tharathep Plienbumrung}
\affiliation{\mbox{Institute for Functional Matter and Quantum Technologies,
		University of Stuttgart, Pfaffenwaldring 57, D-70550 Stuttgart, Germany}}
\affiliation{\mbox{Center for Integrated Quantum Science and Technology,
		University of Stuttgart, Pfaffenwaldring 57, D-70550 Stuttgart, Germany}}
\author{Jean-Baptiste Mor\'ee}
\affiliation{\mbox{RIKEN Center for Emergent Matter Science, Wako, Saitama 351-0198, Japan}}
\author{Andrzej M. Ole\'s$\,$}
\affiliation{\mbox{Max Planck Institute for Solid State Research,
		Heisenbergstrasse 1, D-70569 Stuttgart, Germany} }
\affiliation{\mbox{Institute of Theoretical Physics, Jagiellonian University,
		Profesora Stanis\l{}awa \L{}ojasiewicza 11, PL-30348 Krak\'ow, Poland}}
\author{Maria Daghofer}
\affiliation{\mbox{Institute for Functional Matter and Quantum Technologies,
		University of Stuttgart, Pfaffenwaldring 57, D-70550 Stuttgart, Germany}}
\affiliation{\mbox{Center for Integrated Quantum Science and Technology,
		University of Stuttgart, Pfaffenwaldring 57, D-70550 Stuttgart, Germany}}

\begin{abstract}
We present an effective two-band model for infinite-layer nickelates NdNiO$_2$
that consisting of a  $d$ band centered at Ni site and an interstitial $s$-like band
centered at Nd site. To the large extent of the wave functions, we find
intersite Coulomb interactions to be substantial. We then use the variational cluster
approach together with mean-field theory to investigate magnetic and charge
ordering. While tendencies towards charge modulation are found, they are weak
and might be due to finite-size effects. Magnetic order is determined mostly
by the filling of the $d$ band and hardly affected by including longer-ranged
interactions. For a $d$-band density consistent with density-functional theory, magnetic ordering
vanishes once quantum fluctuations are included to a sufficient spatial
extent. Apart from self-doping,  $d$ and $s$ bands remain largely uncoupled
despite the presence of inter-orbital Coulomb interaction and (small) inter-orbital hopping.
\end{abstract}

\date{\today}

\maketitle


\section{Introduction}
Nickel-based superconductors \cite{Li19,Osa20,Zen22} have
attracted attention mainly due to their potential link to high-$T_C$ cuprate
superconductivity: In both classes of materials, two-dimensional NiO$_2$~\cite{Jia20,Tha21,Hu19,Zha20t}
resp. CuO$_2$ layers form a prominent building block.  Similar to cuprates, which are charge-transfer systems, electron-energy loss spectroscopy supports the picture of 
mixed charge-transfer and Mott-Hubbard characteristics in the nickelate case~\cite{Goo21}. Near the Fermi level,
one finds in both cases a band with pronounced $x^2-y^2$ character, a mostly
two-dimensional dispersion and substantial correlations. Experimentally, strong
antiferromagnetic correlations as well as tendencies towards charge order
connect the two systems.

However, there are also clear differences.  Experimentally, the 'undoped' nickelate  parent compound does not show
cuprate-like long-range antiferromagnetism~\cite{Hay99,Hay03,Ike16,Rathnakaya2025}, but
rather short-range correlations \cite{Fow22,Hsu22} or a glassy
state~\cite{Huan20,Lin22,Ort22}. Theoretically, most methods suggest strong
tendencies towards antiferromagnetism. Strong competition between various charge and
spin instabilities have been found in a weak-coupling analysis~\cite{Lan23},
magnetic excitations calculated in an ordered state closely resemble
experimental spectra of the paramagnetic state~\cite{Lan23,Tha23}. 

There is accordingly some debate on how similar models for nickelates and models
for cuprates can be. In addition to NiO$_2$ layers, which closely resemble
CuO$_2$ layers, rare-earth atoms provide an additional dispersive band
that crosses the Fermi level and hybridizes with nickel states~\cite{Rathnakaya2025}. This band
accepts some electrons in the undoped state and thus proved the 'self doping'
that sets the effective filling of the correlated $x^2-y^ 2$ states. While
this may be this band's most relevant impact~\cite{Kit20}, stronger connections
between the subsystems have also been discussed. For instance, the localized
$x^2-y^2$ states together with itinerant rare-earth bands have been modeled as 
Kondo lattices, which are characterized by antiferromagnetic interactions between
the bands~\cite{Hep20,Wan20p,ZYZ20,Bee21}. Moreover, the dispersive second
band also contains some weight in Ni orbitals ($3z^2-r^2$ as well as $xy$),
which brings into play the possibility of onsite inter-orbital Coulomb and
Hund correlations, as have been treated in a variety of models~\cite{Kre22,Lan23,Tha22}.

Finally, an 
incommensurate  charge-density wave has been reported in layered nickelates, with stripes
parallel to the crystal axes~\cite{Ros22, Tam22}. Its pattern thus differs from charge-density
waves observed in other nickelates and rather resembles doped
cuprates. Coupling to the lattice may here be
involved~\cite{Slo22}, but electronic mechanisms have also been
brought forward. Theoretically, a charge-transfer mechanism in a 17-orbital model has
been put forward as an explanation~\cite{Che23}. Alternatively, the charge
reservoir provided by itinerant states has been shown to enhance tendencies
towards charge order in a Hubbard-ladder~\cite{Shen23}. RIXS
measurement performed in  La$_4$Ni$_3$O$_8$ suggest hybridization of Ni $z^2$
orbitals with rare-earth states to play a substantial role in the CDW~\cite{Shen23b}. A theoretical analysis based on a multi-orbital model has included both onsite Coulomb interactions and non-local correlations and has connected the CDW instability to the Ni-$z^2$ state~\cite{Ste24_CO_nonlocal}.

Here, we derive a two-band model
with longer-ranged interactions via Wannier downfolding and then use the
variational cluster approach to investigate ordered states. One of the bands
has mostly Ni $x^2-y^2$ character, while the other is dominated by rare-earth
states and not centered on Ni sites. While onsite interactions thus
automatically involve only one orbital, longer-ranged Coulomb interactions
also act between different orbitals. We then use the variational cluster approximation (VCA)
\cite{Pot03b,Aic06a,Aic04} to address the ground state of two-band model, see
Sec.~\ref{sec:method}. Section~\ref{sec:res_onsite} then provides results on
the model with purely onsite interactions, where we find antiferromagnetc (\AF) order to be
suppressed once the directly solved cluster becomes large enough. 
In order to treat
inter-site Coulomb interactions, we complement the VCA with a mean-field
embedding similar to that introduced in Ref.~\cite{Aic04} and present results
for nearest-neighbor and longer-ranged interactions in Secs.~\ref{sec:res_NN}
and~\ref{sec:res_lr}, resp.

\section{Model and Method} 

While all models share the $x^2-y^2$ state, they differ in the
description of the remaining systems. In the interest of simplicity, it would
be highly desirable to have a reliable two-band model that might at least be
valid at low doping and small energy scales. The second band can be either an
'axial' $s$ orbital that is not centered on Ni sites~\cite{Adh20}, or another
Ni-bases state~\cite{Kang21}. All these models closely fit
the bands obtained from DFT, so that their band dispersions
are extremely similar. However, wave functions can look very different, which
suggests that electronic correlations might be more sensitive to details of
the model~\cite{Tha22}.

\begin{figure}
  \includegraphics[width=0.9\linewidth]{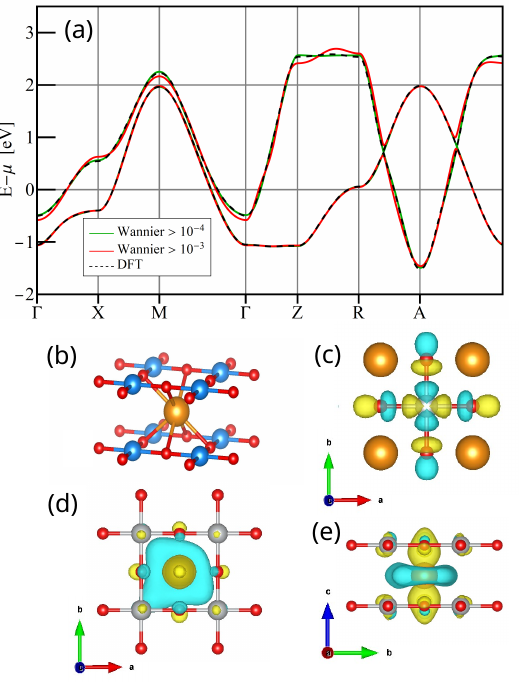}
  \caption{Band dispersion and Wannier orbitals for the two-band model. (a) DFT band structure and
    its Wannier projection. (b) NdNiO$_2$ unit cell, where blue, orange, and
    red denote Ni, Nd, and O atoms. (c) The $x^2-y^2$-like orbital centered
    at an Ni atom. (d) and (e) Top-view  and side-view of the interstitial
    $s$-like orbital centered at an Nd atom.
    \label{fig:bands}}
\end{figure}

More effort has thus been spent on obtaining better models with
as few bands as possible. Substantial onsite inter-orbital correlations between
different Ni orbitals were argued to make three bands a
necessity~\cite{Ste24_CO_nonlocal}. On the other hand, a recent two-band model was obtained and
investigated starting from a tight-binding description of a larger number of
orbitals~\cite{2024arXiv240716042S}. In this paper, we introduce a carefully
derived Wannier-fitted model. We aimed at a model with nearly
real-valued wave functions (a consistency check suggesting good convergence)
that moreover have plausibly symmetric shapes of their real-space wave
functions. (As well as of course a good fit to the DFT band structure.) After
initially finding that we needed three bands to achieve all desiderata, we were
able to prune one band from the three-band description. The wave functions
corresponding to the two remainig orbitals are shown in
Fig.~\ref{fig:bands}(c-e).

Figure~\ref{fig:bands}(a) shows the band structure of NdNiO$_2$ in the vicinity of
the Fermi surface and compares it to the final two-band Wannier fit and its
derived model. Panels (c)-(e) illustrate the wave functions of the two
orbitals and (b) gives the unit cell for comparison.
As illustrated in Fig.~\ref{fig:bands}(c)-(d), the wave function of the $d$ band spreads over
Ni-O bonds, whereas the $s$ orbital wave functions sits mostly 
between the Ni layers, but includes some some apical orbitals from Ni.  While the orbital with $x^2-y^2$ symmetry is robustly obtained
in almost any two-band fit, the shape of the second state can differ
substantially between different fitting procedures, see e.g. the discussion in
Ref.~\cite{Tha22}. 

The effective Wannier band structure is then simplified
by pruning matrix elements smaller than $10^{-3}$, which does not have a
strong impact on the bands, see the colored lines in Fig.~\ref{fig:bands}(a)
The one-particle part that corresponds to the Wannier band structure is then parameterized as 
\begin{align}
\label{eq:ekin}
H_{\rm kin}&=\sum_{i\alpha\sigma}\epsilon^{\alpha}
\dd_{i\alpha\sigma}\dnd_{i\alpha\sigma}
+\sum_{ij\alpha\beta\sigma}{t_{ij}^{\alpha\beta}
  d^{\dagger}_{i\alpha\sigma}d^{}_{j\beta\sigma}},
\end{align}
where $\dnd_{i\alpha\sigma}$ ($\dd_{i\alpha\sigma}$) annihilates (creates) an
electron at site $i$ in orbital $\alpha$ and spin $\sigma$. $\alpha=d$ denotes the Ni-$x^2-y^2$
dominated state of Fig.~\ref{fig:bands}(c) and $\alpha=s$ stands for the state
of Fig.~\ref{fig:bands}(d). Hopping parameters 
$t^{\alpha\beta}_{ij}$ and on-site energies $\epsilon^{\alpha}$ are
given in the Supplemental Material (SM) \cite{suppl}. Let us note here that
$t^{dd}_{ij}$ are strongly two-dimensional, while inter-plane hoppings
dominate  for the $s$ band. Hybridization between the bands is very small.

\begin{figure}
  \includegraphics[width=1\columnwidth]{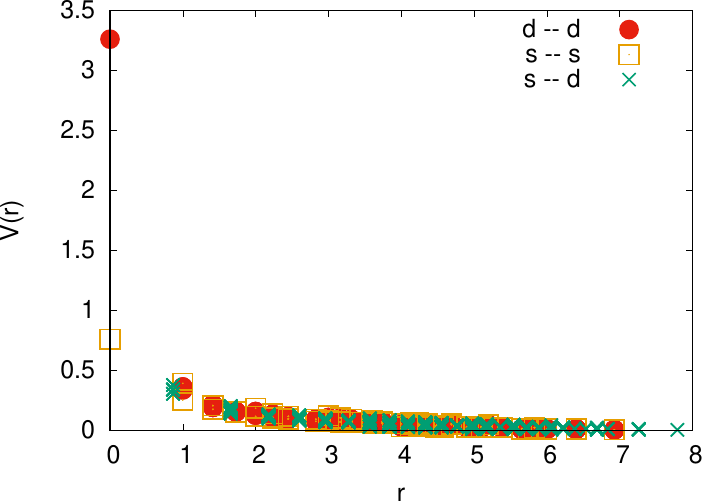}
  \caption{Density-density interactions of the effective two-band
    model. Onsite interactions are automatically intraorbtial and much stronger for the $d$ states (filled circles) than for $s$ states (empty squares). Beyond onsite, interactions between orbitals (crosses) are of similar magnitude as intraorbital interactions.
    \label{fig:U_V_vs_r}}
\end{figure}
Estimates of electronic interactions were obtained using the constrained
random-phase approximation (cRPA)~\cite{Ary04,Ary06}. The cRPA is performed by employing RESPACK code~\cite{Mor22,Nak21}. 
For cRPA calculation, we use 100 bands with a plane-wave cutoff energy of 8 Ry
and an $8\times 8\times 8$ $k$-point grid as in the Wannier projection~\cite{Tha23}. 
The cRPA calculations start from the DFT band structure, which was calculated as described below. We use \texttt{Quantum ESPRESSO}~\cite{QE-2009,QE-2017}, the GGA-PBE functional~\cite{Perdew1996}, and pseudopotentials in which Nd($4f$) electrons are frozen and removed from the valence electrons of the pseudopotential. (These pseudopotentials are the same as those in Ref.~\cite{Nom19}).
We also use a $11 \times 11 \times 11$ $k$-point grid, a plane-wave cutoff energy of $100$ Ry for the wavefunctions, and a $0.002$ Ry Fermi-Dirac smearing.

The dependence of density-density interactions on the distance between orbitals is shown in Fig.~\ref{fig:U_V_vs_r}: As expected,
onsite Coulomb repulsion is considerably weaker for the $s$ than for the $d$
orbital. Inter-orbital interactions are here automatically 'inter site', as the
two orbitals are not centered at the same positions. Such nonlocal interactions i.e.,
inter-site interactions, are non-negligible and
of similar strength for any combination of orbitals ($s$-$s$, $s$-$d$, or
$d$-$d$), see Fig.~\ref{fig:U_V_vs_r}. From the cRPA, it turns out that
Hund's-rule coupling between the $d$ and nearby $s$ orbitals, as well as
between nearby $d$ orbitals,  
is very small ($0.016$ resp. $0.013\;$ eV), so that we neglect them in the rest
of this paper. The Ni-$d$ contribution to the $s$-like orbital of
Fig.~\ref{fig:bands}(d) and (e) is here thus not found to be substantial
enough to induce sizeable Hund's-rule coupling. This procedure results in the  interaction Hamiltonian 
\begin{align}\label{eq:Hint}
  H_{\textrm{int}} = \sum_{i\alpha}U^\alpha
  n_{i\alpha\uparrow}n_{i\alpha\downarrow}
    + \sum_{i,j,\alpha\beta} V^{\alpha\beta}_{ij} n_{i\alpha}n_{j\beta}\;,
\end{align}
with density operator
$n_{i\alpha}=n_{i\alpha\uparrow}+n_{i\alpha\downarrow}$, and couples the
two bands. Here $U^\alpha$ is the onsite Coulomb interaction in orbital $\alpha$: $U^d = 3.261$ eV, $U^s = 0.761$ eV. $V^{\alpha\beta}_{ij}$ is the inter-site Coulomb interaction between orbital $\alpha$ at site $i$ and orbital $\beta$ at site $j$. Note that, $U^d = 3.261$ eV is comparable to the range of values typically obtained for the single-orbital Hamiltonian in cuprates~\cite{Mor22,Mor24Hg1223,Mor24HDE}.

We have recently used the RPA to investigate magnetic instabilities of the
model restricted to onsite (and thus intra-orbital)
interactions~\cite{Tha23} and found an instability towards $G$-type
magnetic order (closely competing with $C$-type) and magnetic excitations
consistent with a more complex model that is also used onsite
interactions~\cite{Lan23}. When including longer-ranged interactions, however,
a larger number of potential orderings emerge already at quite weak
interactions resp. high temperatures. While this supports the notion that the
two-band model with long-range interactions captures some of the competing
tendencies observed in more complex models, it also suggests that the
weak-coupling approach becomes ineffective. We consequently use the
variational cluster approach.

\subsection{Variational Cluster Approximation} \label{sec:method}
\begin{figure}
     \centering
    \includegraphics[width=1\linewidth]{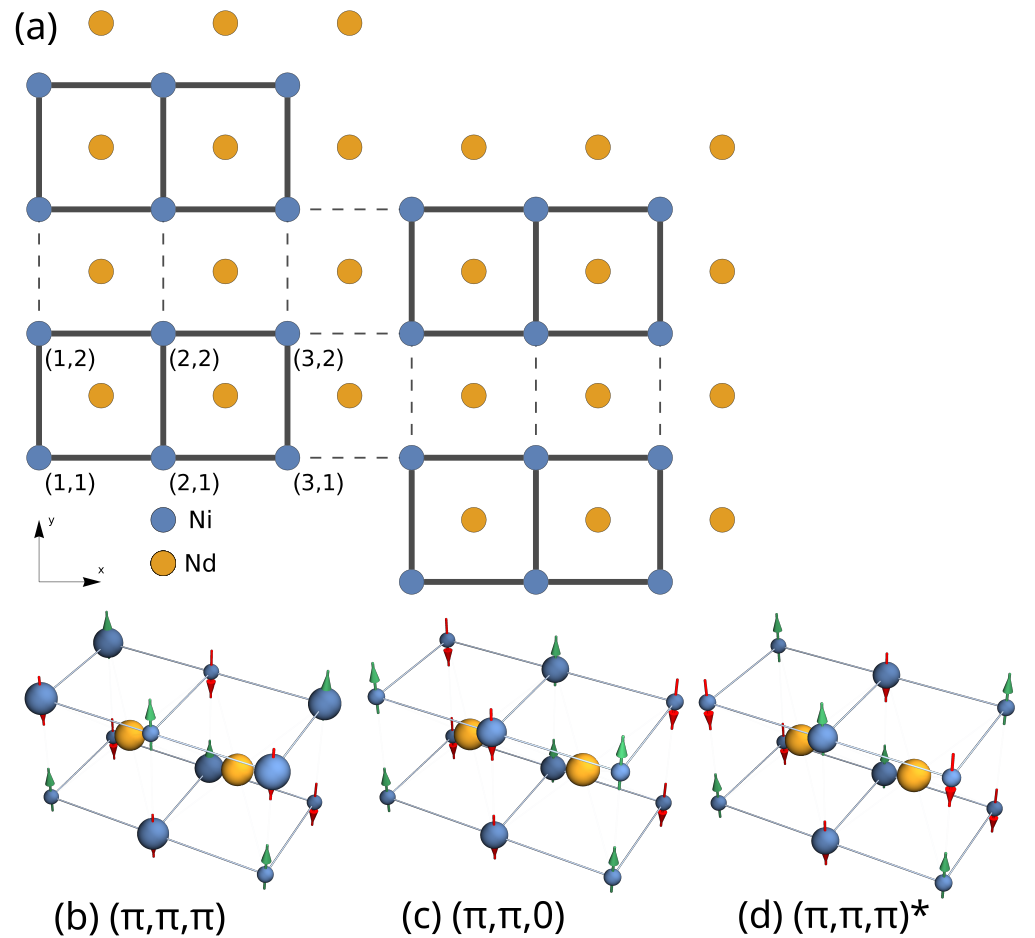}
    \caption{Representative cluster. (a) The lattice system in the xy plane is constructed by connecting the $3\times 2\times 1$ cluster (solid line) to other identical clusters. (b)-(d) Charge and Spin orientation in the three-dimensional cluster. The labels $(\pi,\pi,\pi)$, $(\pi,\pi,0)$, and $(\pi,\pi,\pi)^\ast$ refer to magnetic ordering vector $\vec{Q}$ (see main text). (b) $G$-type AF with alternating charge in the out-of-plane axis. (c) $C$-type \AF~with uniform charge in the out-of-plane axis. (d) $G$-type \AF~with uniform charge in the out-of-plane axis.}
    \label{fig:cluster}
\end{figure}
To study the two-band Hamiltonian, we employ the variational cluster approximation~\cite{Pot03b,Aic04,Aic06a}. The method can be viewed as an extension of
cluster perturbation theory (CPT) \cite{Sen00,Sen02}, or a specific case of
self-energy functional theory (SFT) \cite{Pot03a}. The underlying idea of VCA relies on dividing the lattice system into smaller
clusters, which is exactly solvable e.g., Fig.~\ref{fig:cluster}(a), and then calculate the grand canonical potential
$\Omega$ of lattice system via cluster self-energy $\Sigma_{\text{Cl}}$ at a
stationary point. The grand canonical potential of the lattice system can
be written as 
\begin{equation}
	\Omega = \Omega_\text{Cl} + Tr\ln{[G^{-1}_0 - \Sigma_{\text{Cl}}]^{-1}} - Tr\ln{(-G_\text{Cl})},
	\label{eq:omega}
\end{equation}
where $\Omega_\text{Cl}$, $G_\text{Cl}$ are grand canonical potential and Green's
function of the cluster. $G^{-1}_0$ is the non-interacting lattice
Green's function. The lattice Green's function in a Dyson-like form
is 
\begin{align}\label{eq:green_CPT}
G = (G^{-1}_0 - \Sigma)^{-1}\;, 
\end{align}
with the approximation $\Sigma \approx \Sigma_{\text{Cl}}$. 

According to SFT, $\Sigma_{\text{Cl}}$ can be optimized by varying
the one-particle terms $t'$ of the cluster Hamiltonian, e.g., chemical potential,
crystal field-splitting, fictitious symmetry breaking field etc. The optimal
value of the parameters is determined from the stationary point of
(\ref{eq:omega}) i.e., $\frac{\partial{\Omega}}{\partial{t'}} =
0$~\cite{Pot03a,Aic06b}. Note that when calculating lattice quantities
(Green's function, grand potential or densities) the physical Hamiltonian of interest is
used. As an example, a self-energy calculated with a staggered magnetic field
might optimize the grand potential of a fully symmetric Hamiltonian, which
would then indicate spontaneous symmetry breaking.

In principle, varying all possible single-particle operators of $H_{\text{Cl}}$ is
desired. This requires searching for a stationary point of all varied
operators, which in practice makes optimization intractable. We consequently focus here on a
fictitious chemical potential 
\begin{align}
  H_{{\mu'}} = \mu'\sum_{i,\alpha} n_{i,\alpha}
\end{align}
needed for thermodynamic
consistency~\cite{Aic05}
and a staggered magnetic field
\begin{align}\label{eq:mag}
  H_{h'} = h'\sum_{i,\alpha} e^{i\Vec{Q}\cdot\Vec{R}_{i}}(n_{i,\alpha,\uparrow}-n_{i,\alpha,\downarrow}) = h'M.
\end{align}
In particular, we use ordering vector $\Vec{Q} = (\pi,\pi,0)$ for $C$-type AF and $\Vec{Q} = (\pi,\pi,\pi)$ for $G$-type AF. Parameter $h'$ is varied to optimize the grand potential and $M$ is the staggered magnetization serving as AF order parameter.
Additionally, we use a Legendre transform from the grand potential to the free
energy to obtain results at a fixed particle
number rather than for a fixed chemical potential~\cite{Bal10}. This allows us to compare
energies of different ordered phases that may require different chemical potentials.

To extend the VCA beyond onsite interactions, we employ a
mean-field decoupling of inter-cluster interactions. Such an approach has been
used successfully in investigating charge order in extended Hubbard
models~\cite{Aic04, Dag14}. In this case and especially for small clusters, the
symmetry breaking is provided mainly by the mean-field parameters: the VCA
calculation is performed for each set of mean-field parameters and new
parameters are obtained from results until self consistency is reached.

After convergence and for optimal parameters, we obtain physical quantities
like orbital-resolved densities and staggered magnetization
(\ref{eq:mag}). Additionally, we obtain the one-particle spectral density from
the CPT Green's function (\ref{eq:green_CPT}). 

\subsection{Magnetic and charge patterns}

We perform our calculation mainly on a $3\times2$-unit cell; see Fig.~\ref{fig:cluster}(a). This is suggested by the observation of incommensurate CDW in the infinite-layer nickelates~\cite{Tam22,Ros22}. Each unit cell consists of a Ni-centered $d$
orbital and an interstitial-$s$ orbital, and the clusters are stacked to permit alternating in-plane magnetic order in addition to a CDW. Depending
on the stacking perpendicular to the plane, three phases with in-plane
antiferromagnetism are accessible, see Fig.~\ref{fig:cluster}(b)-(d): Stacking vector $(0,0,1)$ (in unit cells)
corresponds to $C$-type magnetic ordering vector $\vec{Q}= (\pi,\pi,0)$, where sites with equivalent
charge order are on top of each other (Ferro-charge). Stacking vector $(1,0,1)$ gives
$G$-type $\vec{Q} = (\pi,\pi,\pi)$ antiferromagnetism with a charge pattern shifted by one (out
of three) unit cells along $x$ (Anti-ferro charge). Finally, $(0,1,1)$ combines $\vec{Q} = (\pi,\pi,\pi)$
antiferromagnetism with a charge pattern shifted along $y$, which implies for
a perfect stripe pattern that equivalent sites are on top of each other
(Ferro-charge again). This
latter phase is denoted by $\vec{Q} = (\pi,\pi,\pi)^{\ast}$. In addition, we
find that the \AF~patterns with in-plane stripes, i.e. $\vec{Q} = (\pi,0,0)$ and $\vec{Q} =
(\pi,0,\pi)$ are not favored over the paramagnetic state.  (They are thus not
going to be discussed any further.)

\section{Results}

To gauge the impact of long-ranged interactions, we 
compare the model with intersite interactions to one with purely onsite interactions. In both
cases, we investigate charge and magnetic order as well as their interplay. We focus here on quarter-filling, i.e., one electron per site.
To correct for
double-counting of interactions, we adjust the crystal-field splitting
$\Delta= \epsilon^s -\epsilon^d$, see (\ref{eq:ekin}), to obtain on average
the same densities ion $s$- and $d$-bands as in DFT. Since correct treatment
of double-counting is a non-trivial issue and since different $d$-band
occupations from 7$\%$-17$\%$ have been reported in the
literature~\cite{Gu20,Bot20,Nom19}, we will present results for
several choices of the crystal field to assess how strongly 
corrections might affect the physical picture.

\subsection{Onsite interactions}\label{sec:res_onsite}


\begin{table}
	\begin{tabular}{ |c|c|c|c|c| } 
		\hline
		  n$_d$ & n$_s$ & n$_{tot}$ &  $\vec{Q}$ & $|M|$ \\
		\hline 
		0.711 (0.704) & 0.293 (0.293) & 1.004 (0.996) & ($\pi,\pi,0$) & 0.278\\ 
		0.846 (0.847) & 0.154 (0.153) & 1.000 (1.000) & ($\pi,\pi,0$) & 0.287\\
        0.881 (0.889) & 0.119 (0.114) & 1.000 (1.004) &  ($\pi,\pi,0$) & 0.496\\
        0.926 (0.938) & 0.073 (0.062) & 0.999 (1.000) & ($\pi,\pi,\pi$) & 0.634 \\
		\hline
	\end{tabular}
	\caption{Ground state profile in the paramagnetic state for various
          occupations on the $s$-orbital $n_s(h'=0)$. The values in parentheses correspond to those in the AF state $n_s(h'_\textrm{opt})$. }
	\label{tab:onsite}
\end{table}

\begin{figure}
    \centering
    \includegraphics[width=0.9\linewidth]{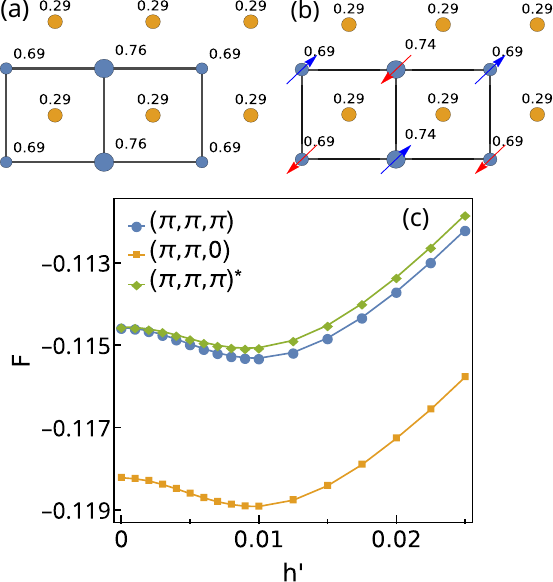}
    \caption{VCA results for 'bare' model parameters with only onsite interactions. (a) Charge pattern at paramagnetic state $h'=0$.
    (b) for C-type AF state at $h'_\text{opt}=0.009$.
    (c) Free energy
    as a function of the fictitious magnetic ordering field $h'$.} 
    \label{fig:undoped}
\end{figure}

We first investigate the system in the onsite-only case, i.e., by setting
$V^{\alpha\beta}_{ij}=0$ in (\ref{eq:Hint}) for $i\neq j$, which automatically
also removes any inter-orbital interaction $\alpha\neq
\beta$. Searching for stationary point w.r.t. $\mu$, $\mu'$, and $h'$, we find
long-range magnetic ordering that can go together with spatial charge
modulation.

In this
'bare' model parameters, the stronger Coulomb repulsion within the $d$ band
pushes electrons into the $s$ band. This effect is rather strong, giving almost 30$\;\%$ of the electrons occupy $s$ states. The $d$ band is then
heavily self-doped and shows charge modulation, see
Fig.~\ref{fig:undoped}(a), which is slightly weakened for the optimal magnetic ordering field $h'=0.01$, see
Fig.~\ref{fig:undoped}(b). The free energy in Fig.~\ref{fig:undoped}(c) shows
a small energy gain for all magnetic patterns, favoring $C$-type order. 

\begin{figure}
  \includegraphics[width=0.90\columnwidth]{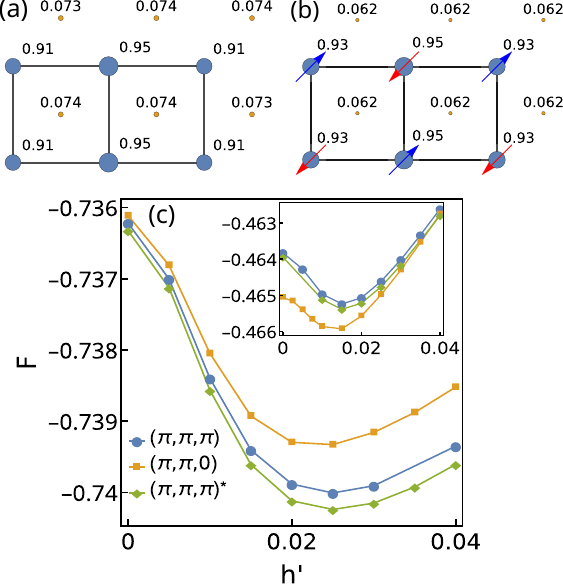}
  \caption{VCA results for adjusted crystal field to account for double\jbm{-}counting. (a)
    The charge pattern at paramagnetic state $h'=0$ 
    (b) for G-type \AF~state at $h'=0.025$, both for a $d$-orbital density of
    $n_{d}\approx 0.93$.
    (c) Free energy
    as a function of the fictitious magnetic ordering field $h'$. Inset shows
    the same for a crystal field giving increased self doping $n_d =
    0.89$.\label{fig:undoped_adj}
  }
\end{figure}

However, there is no symmetry of $d$-and $s$-states canceling double    
counting, so the observed change in orbital occupations may be due to
the double-counting of Coulomb interactions. Following~\cite{V_I_Anisimov_1997}, a
correction of the $d$-level onsite potential $\epsilon^{d}$ by $\approx
\tfrac{U^{d}}{2}$ is expected. An onsite-energy correction by $\approx
1.1\;\textrm{eV}$ would be expected, which yields $n_s = 0.07$. This is not far
off the DFT value of $n_s = 
0.09\;\textrm{or}\;0.1$, which can be reached with a correction of $0.8\;\textrm{eV}$. 

Figure~\ref{fig:undoped_adj}(a) shows electronic densities in the paramagnetic
state and  $n_s = 0.07$, obtained with a double-counting correction of
$\tfrac{U^{d}}{2}=1.1$ eV. Similar to the 'bare' model, larger densities are
in the center sites of the cluster, but the charge modulation has only about
half the size. Densities in the optimal AF state are shown in
Fig.~\ref{fig:undoped_adj}(b) and show a further suppression of charge
modulation. Fig.~\ref{fig:undoped_adj}(c) gives the grand potential depending on
the fictitious ordering field $h'$ for various AF patterns, with $G$-type AF coming
out as the ground state. The system thus largely
recovers the behavior of a weakly doped single-band Hubbard model: robust AF
order without (significant) charge modulation, as seen in
Fig.~\ref{fig:undoped_adj}(b). Slightly increasing the occupation in the $s$-band to $n_s = 0.12$,
leads to the transition from $G$-type to $C$-type \AF~ground state, see inset of
Fig.~\ref{fig:undoped_adj}(c), but to otherwise consistent results.

Densities and preferred ordering are summarized in Table~\ref{tab:onsite} for
several values of $n_s$. When $n_s$ is less than $\approx 10\%$, 
$G$-type AF with ordering vector [$\vec{Q} = (\pi,\pi,\pi)$ or
$(\pi,\pi,\pi)^*$ with almost the same energy] becomes the ground state. $C$-type AF is stable for
higher $n_s$. These three patterns only differ in the $c$-direction where
the $d$-band has almost no dispersion. CDW with slight charge modulation can here be found with purely onsite
interactions and coexiting with \AF~order for small self doping resp. in the
paramagnetic state with larger self doping. However, the tendency to the charge order requires substantial
(self-)doping of the $d$ orbital. 
As the small size of the directly solved cluster is expected
to further enhance ordering~\cite{Dah04,Aic04}, we will examine the finite-size effect in the
next section, before discussing the impact of intersite interactions.

\subsection{Impact of cluster size: absence of magnetic order}\label{sec:res_finitesize}

\begin{figure}
    \centering
    \includegraphics[width=1\columnwidth]{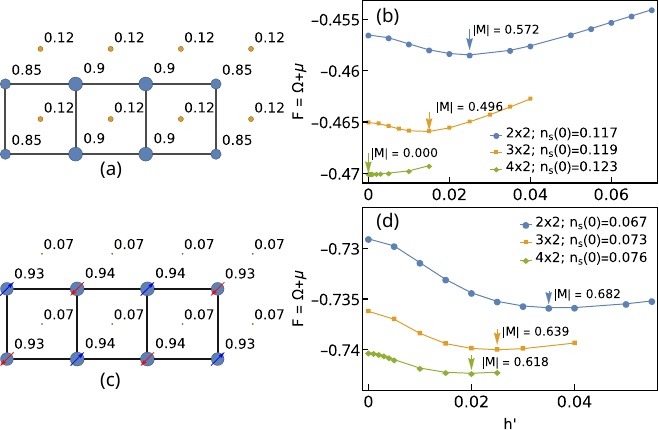}
    \caption{Finite-size scaling. (a) Charge pattern for $n_s(h'=0) =
      0.12$. (b) Free energy as a function of the staggered field $h'$ for
      several cluster sizes at $n_s(h'=0) = 0.12$. (c)-(d) Similar to (a)-(b)
      but $n_s(h'=0) = 0.07$. Arrows in (b)\&(d) indicate the optimal point
      $h'_\text{opt}$.} 
    \label{fig:4x2}
\end{figure}

\begin{figure*}
  \includegraphics[width=0.9\linewidth]{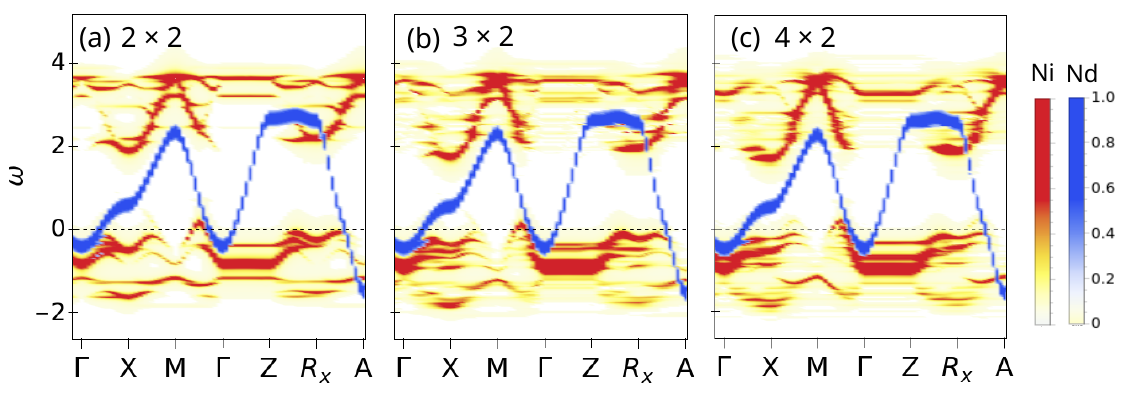}
   \caption{Single-particle spectral function for the onsite-only case at $n_{s}\approx 0.11$ and optimal Weiss field $h' = h'_\text{opt}$: (a) $2\times 2$ cluster in \AF~state. (2) $3\times2$ cluster in \AF~state. (c) $4\times2$ cluster in paramagnetic state, where $h'_\text{opt} = 0$. 
    \label{fig:spec_V0}}
\end{figure*}

To verify the stability of the \AF~solutions, we repeat the above
calculations on a larger ladder i.e., a $4\times2$ cluster, which
is shown  in Fig.~\ref{fig:4x2}. Since these calculations push the limits of
the available computing power, we were not able to combine them with a
self-consistent determination of the charge-order pattern.

For a self-doping level of $n_{d} \approx 0.88$,
i.e. $n_s(h'=0) \approx 12\;\%$,  $C$-type \AF~order disappears for the larger
cluster. The optimal $h'$ is reduced from 0.025 ($2\times 2$ sites) to 0.015
($3\times 2$) and then $0$ ($4\times 2$). The staggered magnetizations shrink from $M=0.57$ via $M=0.5$ to
$M=0$, where $M=1$ would correspond to perfect order. A slight charge
modulation with larger density in the middle persists, again only on Ni sites,
of a similar size as above (difference of $0.04$).

Figure~\ref{fig:spec_V0} shows the spectral density for self-doping
$n_{d} \approx 0.89$ resp. $n_{s}\approx 0.11$, i.e., close to DFT values, depending on the size of the
directly solved clusters. The small hybridization included in the hopping
elements is clearly not strong enough to induce sizable band mixing, so that the
$d$-states largely reproduce results for a single-band Hubbard
model~\cite{Sen05,Arr09,Aic04,Aic06a}. Similarly, the more dispersive and
clearly three-dimensional $s$-band only slightly hybridizes (around the
$\Gamma$ point) with the $d$ states. \AF~order selected for the smaller clusters leads to
band folding of the $d$ states, with the $s$ states being
unaffected and in fact showing practically no signatures of correlations. In the spectral density of the $4\times 2$ cluster, which does not
show long-range \AF~order, these features are accordingly much weaker: The
$d$ band going down to $\approx 1\;\textrm{eV}$ around the $\Gamma$-point is
mirrored onto $M$ in (a) and (b), but not in (c). Only close to the Fermi
level between $M$ and $\Gamma$, slight band folding provides a hole pocket.

In the case of a smaller self-doping $n_{d} \approx 0.93$, i.e. $n_s(h'=0)
\approx 0.07\;\%$, ordering field and ordered moment are also reduced, with
$M=0.68$ ($2\times 2$), $M=0.64$ ($3\times 2$), and  $M=0.61$ ($4\times
2$). As shown in Figs.~\ref{fig:4x2}(c)-(d), the magnetic ordering survives here, while charge modulation is
suppressed by it. We thus find a competition between an \AF~state without
charge order and a nonmagnetic one with (slight) charge modulation, triggered
by self-doping, reminiscent of the scenario discussed in
Ref.~\cite{Che23}. However, our results are consistent with either or both
order types vanishing in the thermodynamic limit. 

\subsection{Nearest-Neighbor interactions}\label{sec:res_NN}
\begin{figure*}
    \centering
    \includegraphics[width=0.9\linewidth]{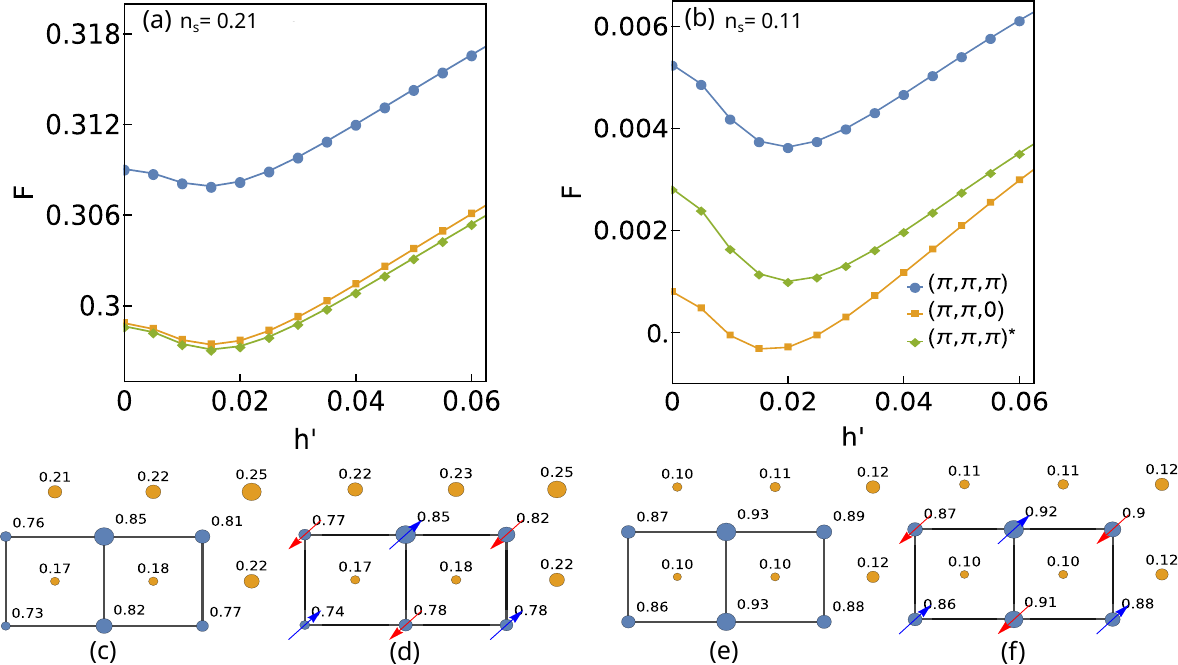}
    \caption{Magnetic order and charge pattern for the model including NN
      interactions. (a) Free energy depending on magnetic ordering field for the 'bare' model parameters with $n_d
      = 0.786$. (b) Analogously for an adjusted crystal field giving $n_d =
      0.892$, close to DFT densities. (c)-(d) Charge patterns for $n_s =
      0.20$: $h' = 0$ in (c), and $h' = 0.015$ in (d). (e)-(f) Charge patterns
      for $n_s = 0.11$: $h' = 0$ in (e), and $h' = 0.015$ in (f) 
    \label{fig:CDW_NN}
  }
\end{figure*}

We further investigate the impact of longer-ranged and inter-orbital interactions of
CDW and \AF~order. As can be seen in Fig.~\ref{fig:U_V_vs_r}(a), all inter-site interactions  are
considerably weaker than onsite ones. However, each Ni-centered $d$ orbital is surrounded by eight
nearest-neighbor interstitial-$s$ orbitals as well as six nearest-neighbor $d$
sites, see Fig.~\ref{fig:U_V_vs_r}(b). The intersite Coulomb energy  
$V_{ij}^{\alpha\beta}$ in (\ref{eq:Hint}), which is proportional to the number
of these neighbors, can be significantly enhanced such that it competes with
the onsite Coulomb energy.  
We thus include the closest of these longer-ranged interactions, namely the inter-orbital
interactions $V^{sd}_{ij}$ between a Ni site and its closest Nd sites as well as nearest-neighbor intraorbital interactions $V^{ss}_{ij}$ and $V^{dd}_{ij}$. 

Including longer-ranged interactions, which act on both
orbitals, actually reduces the double-counting problem, the bare crystal
fields now yield $n_{s}\approx 20\;\%$. We present the corresponding results
and compare them to those obtained for crystal fields adjusted to give  $n_{s}\approx
11\;\%$, close the DFT value. As for purely onsite interactions, magnetic
ordering is stronger for smaller self doping, see Figs.~\ref{fig:CDW_NN}(a) and
(b). 

The combined VCA--mean--field calculations give both magnetic and charge
order. For the larger self doping, the charge patterns are shown in
Fig.~\ref{fig:CDW_NN}(c) without and (d) with magnetic
ordering. Charge modulation without magnetic
ordering is mostly along one direction, i.e., close to $(\pi/3,0)$ in
plane, but also has a modulation along the second direction. In the presence
of magnetism, it becomes more two-dimensional, so that larger 
clusters would be needed. For the adjusted crystal
fields, charge modulations are weaker, comparable to the model with purely
onsite interactions.  

Magnetic order is again confined to the $d$ orbitals, but charge modulation
slightly spills over into the $s$ orbitals.  RIXS experiments have observed
charge modulations to be mostly on Ni orbitals, with weaker contributions from
other states, in that case oxygen~\cite{Tam22,Ros22}. In our model, modulation
of $s$-orbital density can be understood from the
inter-site interactions between $s$ and $d$ orbitals. Comparing the inset of
Fig.~\ref{fig:undoped_adj} and Fig.~\ref{fig:CDW_NN}(b) shows that NN
interactions enhance the energy difference between $(\pi,\pi,\pi)$ and
$(\pi,\pi,\pi)^{\ast}$ orderings. In the second case, Ni sites with larger charges are on top of each
other. Such a pattern can be stabilized by $d$-$s$ interactions, because the
larger $d$-charges are then surrounded by those $s$ orbitals with smallest
charge. Table~\ref{tab:nn} provides electron densities and their respective AF ground state. 

\begin{table}
	\begin{tabular}{ |c|c|c|c|c| } 
		\hline
		  n$_d$ & n$_s$ & n$_{tot}$ &  $\vec{Q}$ & $|M|$\\
		\hline 
		0.786 (0.791) & 0.213 (0.213) & 0.999 (1.001) &  $(\pi,\pi,\pi)^*$ & 0.413 \\
		0.847 (0.848) & 0.156 (0.155) & 1.003 (1.003) & ($\pi,\pi,0$) & 0.441 \\
        0.892 (0.891) & 0.109 (0.109) & 1.001 (1.000) &  ($\pi,\pi,0$) & 0.474 \\
        0.898 (0.902) & 0.102 (0.099) & 1.000 (1.001) &  ($\pi,\pi,0$) & 0.478 \\
		\hline
	\end{tabular}
	\caption{Ground state profile for various $n_s$ when intersite Coulomb interactions are included. The values in parentheses correspond to those in the AF state $n_s(h'_\textrm{opt})$.}
	\label{tab:nn}
\end{table}

\begin{figure}
    \centering
    \includegraphics[width=0.9\columnwidth]{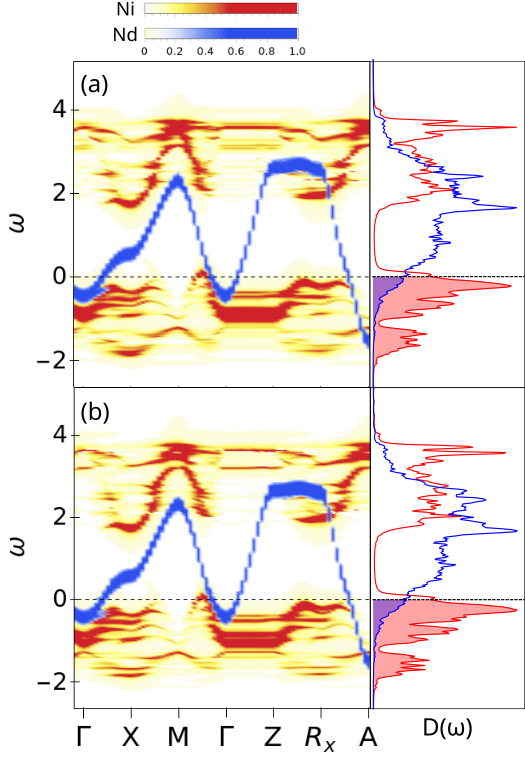}
    \caption{Spectral function and density of states (DOS) in AF solutions for n$_s$ = 0.12 (onsite), and n$_s$ = 0.11 (including NN interactions). (a) only onsite interactions. (b) including NN interactions.
    \label{fig:dos}
  }
\end{figure}

Figure~\ref{fig:dos} shows the single-particle spectral functions
of the two-band model in the AF state, both for purely onsite interactions and
including NN Coulomb terms. Differences are hardly noticeable, and in fact
hard to resolve from numerical errors. 

\subsection{Beyond nearest-neighbor interactions}\label{sec:res_lr}
\begin{figure}
    \centering
    \includegraphics[width=1\columnwidth]{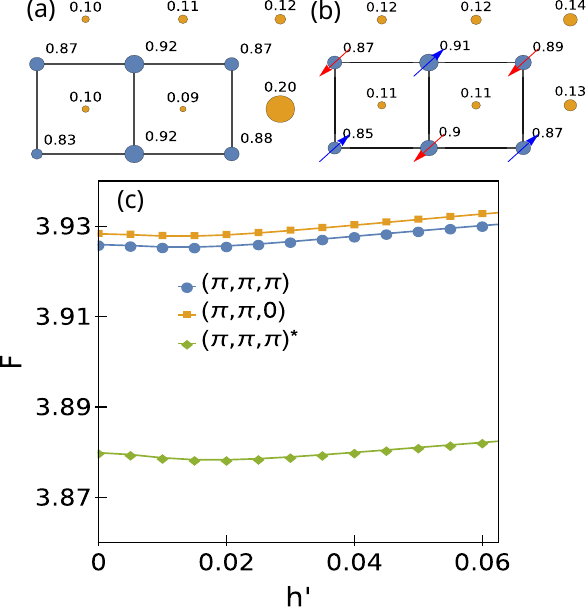}
     \caption{ VCA results for 'bare' model parameters with longer-ranged interactions. (a) Charge pattern at paramagnetic state $h'=0$.
    (b) For G-type \AF~(Ferro-charge), the state is $h'=0.02$.
    (c) Free energy $F$
    as a function of the fictitious magnetic ordering field $h'$. \label{fig:long}}
\end{figure}

For consistency, we further include all density-density interactions
that can be reached within the $3\times2$ cluster, again using the
mean-field decoupling for the bonds connecting clusters. In that case, the
densities found in $d$- and $s$-orbitals are $n_{d}= 0.88$, and $n_{s}= 0.12$,
i.e. quite close to DFT values. We accordingly do not use any double-counting correction here. 

The charge patterns obtained self-consistently for the state without and with
\AF~order are shown Figs.~\ref{fig:long}(a) and~(b). Similar to the previous
cases, we find a slightly larger charge at the center sites of each
leg. Magnetic ordering slightly reduces 
it, and preferred stacking is $(\pi,\pi,\pi)^{\ast}$, i.e., $G$-type \AF
~ferro-charge along the $z$-axis. Without \AF~order, there is now a pronounced
charge imbalance wihtin the $s$-states, which may be the reason why the $(\pi,\pi,\pi)^{\ast}$-AF pattern is now
preferred, see Fig.~\ref{fig:long}(c): it avoids stacking large
$s$-densities on top of each other. However, this may be a finite-size
effect. Robust results thus remain largely unchanged compared
to only NN interactions (and not very different from those with purely onsite
Coulomb repulsion).

\section{Summary and conclusions}

We have investigated the charge and spin orders in a two-band model for
NdNiO$_2$. The derived model is well converged with the wave functions in
both bands having some Ni-orbital content. Nevertheless, the effective
Hund's-rule coupling has practically vanished, while inter-site Coulomb
interactions are substantial.  

To address the ground-state properties, we employ the VCA, focusing on a $3\times2$ cluster. In
the magnetic sector, we find the \AF~orders, $G$-type, and $C$-type, when the
$d$-band is not self-doped too far away from half-filling, $n_s = 1-n_d \approx 7-15\%$. 
However, these \AF~orders are absent when considering the larger $(4\times2)$
cluster at the self-doping $n_s \approx 11\%$, which is close to that obtained
from the DFT. This corroborates the picture of long-range \AF~order being
suppressed by self-doping, without requiring substantial
competing charge order. 

In the charge sector, we find some tendencies towards stripy charge order with a periodicity of
three sites, i.e. close the incommensurate ordering vectors reported, but
charge modulation is rather weak and does leave signatures in the
one-particle spectral density. Therefore, we conclude that while tendencies
towards charge order formation might be present, CDW formation would likely
need additional triggers beyond inter-site Coulomb interactions. 

In addition to small hybridization in the hopping elements, the two orbitals
are, in our model, connected by the long-ranged Coulomb interactions. We find
that if a CDW is formed, the $s$ band may tip the balance between phases that
differ in their stacking along the $z$-direction, as $s$-sites with lower
charge can stabilize ferro-charge ordering of Ni sites.

\section*{Acknowledgment}
The calculations in this article are supported by the state of Baden-W\"urttemberg through bwHPC and the German Research Foundation (DFG) through grant no INST 40/575-1 FUGG (JUSTUS 2 cluster). T.P. acknowledges the Development and Promotion of Science and Technology Talents Project (DPST).
J.-B.M. acknowledges support by the RIKEN TRIP initiative (RIKEN Quantum, Advanced General Intelligence for Science Program, Many-body Electron Systems).

\bibliography{bib_v6}

\end{document}